\newcommand{\pa}{\partial}
\newcommand{\be}{\begin{equation}}
\newcommand{\ee}{\end{equation}}
\newcommand{\bea}{\begin{eqnarray}}
\newcommand{\eea}{\end{eqnarray}}

\newcommand{\nn}{\nonumber}

\newcommand{\bt}[1]{{\bar t}}

\documentclass[12pt]{article}

\input epsf
\usepackage{amsmath,amsfonts,epsfig,color,latexsym}
\usepackage{amssymb}
\usepackage{amsfonts}
\usepackage{amssymb}
\def \a {\alpha}

\def\s{\sigma}

\def\vep{\varepsilon}

\def \vp {\varphi}

\def \s{\sigma}
\def \pa{\partial}

\def \ha {{1 \over 2}}

\def\gd{\nu }

\renewcommand{\thefootnote}{\fnsymbol{footnote}}

\def\appendix#1{
  \addtocounter{section}{1}
  \setcounter{equation}{0}
  \renewcommand{\thesection}{\Alph{section}}
  \section*{Appendix \thesection\protect\indent \parbox[t]{11.15cm}
  {#1} }
  \addcontentsline{toc}{section}{Appendix \thesection\ \ \ #1}
  }

\def\be{\begin{equation}}
\def\ee{\end{equation}}

\def \bi{\bibitem}

\def \ha {{1 \over 2}}
\def \td {\tilde}

\begin{document}


\null\vskip-24pt 
\hfill {\tt hep-th/0508125}
\vskip0.2truecm
\begin{center}
\vskip 0.2truecm {\Large\bf
String spectrum of curved string backgrounds\\
\vskip 0.2truecm
obtained by T-duality and shifts of polar angles
}
\\
\vskip 0.7truecm
{\bf Jorge G. Russo}\\
\vskip 0.4truecm

{\it 
Instituci\' o Catalana de Recerca i Estudis Avan\c{c}ats (ICREA),\\
Departament ECM,
Facultat de F\'\i sica, Universitat de Barcelona,
 Spain} 
\end{center}
\vskip 0.2truecm 
\noindent\centerline{\bf Abstract}
A class of exactly solvable string models can be obtained
by starting with flat space and
combining T-duality and shifts of  angular coordinates of several
polar planes.
The models are the analog of the  Lunin-Maldacena
$\beta $-deformation of the $AdS_5\times S^5$ type IIB string
background, which is dual
to a Leigh-Strassler deformation of ${\cal N}=4$ Super Yang-Mills Theory.
We determine the complete physical string spectrum for two string models obtained in this way, by
explicitly solving the string equations and quantizing in terms of
free 
creation and annihilation operators.
We also show that the 3-parameter $(b_1,\, b_2,\, b_3)$ model, 
obtained by three independent TsT transformations,
has tachyons in some regions of the parameter space.

\newpage

\renewcommand{\thefootnote}{\arabic{footnote}}
\setcounter{footnote}{0}

\section{Introduction}

Conformal theories representing strings in curved backgrounds are in
general extremely complicated and the physical string spectrum is
known
only in a few cases.
One of these cases is the class of string backgrounds obtained by a
sequence of T-duality transformations 
and shifts of periodic coordinates involving other periodic
coordinates
of different periods \cite{RT}.  Due to the shifts, the
new conformal field theories are not equivalent to the original flat-space starting point. 
By construction, the resulting model is nevertheless an exact
conformal field theory, to all orders
in the $\alpha '$ expansion.
The models, being completely solvable,
were used to test 
physical aspects of string propagation in curved spacetime, including
closed superstrings in magnetic fields \cite{magnetic},
supersymmetry breaking and 
 closed string tachyons \cite{magnetic,taka,flux}, 
 D branes in
 magnetic fields \cite{taka2}, decay of type 0 string vacuum
 \cite{costa,RT01}, spacetime singularities \cite{RT}, strings in plane
wave backgrounds \cite{blau,hashi} and closed time-like
 curves \cite{fiol}.

Finding new conformal $\sigma $ models of strings in curved
backgrounds where the string spectrum can be found exactly
is of obvious interest.
New solvable models have been recently introduced by Lunin and
Maldacena in \cite{LM}.
The models are constructed by applying a T duality
transformation, then a 
shift of a periodic coordinate, and then another T-duality
transformation. These transformations were generally referred as 
TsT transformations in \cite{frt}, and here we will adopt this name.
In contrast to \cite{RT}, where the shift
involves an $S^1$ coordinate, in \cite{LM} 
the shift only involves  polar angles. This novel application
of TsT transformations gives rise to new
exactly solvable string models which have not been studied before.

The main motivation for the study of these models is that they are
closely related to the  analogous
deformation
of $AdS_5\times S^5$ that yields a background $(AdS_5\times
 S^5)_\beta$, 
which is the
 supergravity dual of  the Leigh-Strassler $\beta $-deformation of ${\cal N}=4$
 Super Yang-Mills Theory to a ${\cal N}=1$ superconformal gauge theory
\cite{LEST} (some recent interesting works on the 
$(AdS_5\times S^5)_\beta$ string theory can be found in
\cite{NIPR,frolov,nunez,beisert,koch,mat,bobev,freedman,penati,frt2,Ahn,kuzenko}).
The Lunin-Maldacena deformation applies to the $S^5$ part of the
 space.
The 5-sphere can be represented by three complex planes with the
 restriction
$z_1z_1^*+ z_2z_2^*+ z_3z_3^*=R^2$, where $R$ is the
 radius of the sphere (related to the 't Hooft coupling $\lambda=g_{\rm YM}^2N$ by
$R^2=\sqrt{\lambda }\, \a' $). The present models are obtained  
by a similar deformation applied on a 
flat space $(z_1,z_2,z_3)$.

This paper is organized as follows. In section 2 we consider 
the simplest model involving two complex planes, and find the mass spectrum
of quantum superstring states. 
A model involving three complex planes related to the 
Lunin-Maldacena deformation is then considered in section 3.1.
In section 3.2 we consider a model which is the analog of the
three-parameter deformation  of  $AdS_5\times S^5$  
introduced by Frolov \cite{frolov} and further studied in \cite{frt2}. 
We find tachyon states in some regions of the parameter
space. Finally, section 4 contains some remarks on
the string spectrum in $\beta$-deformed $AdS$ backgrounds.

\setcounter{equation}{0}

\section{String model I: TsT on two polar planes}

This model was introduced in \cite{LM} to illustrate TsT
transformations in a simple setting. The starting Lagrangian is
\be
L=\pa_+x_\mu \pa_-x^\mu+
\pa _+ r_1\pa _- r_1+r_1^2\pa _+\vp_1\pa _-\vp_1 +
\pa _+ r_2\pa _- r_2+r_2^2\pa _+\vp_2\pa _-\vp_2\ ,\ \ \ \ \ 
\mu=0,1,...,5
\ .
\label{aaa}
\ee
We use the notation $\s_\pm=\tau\pm\s $.
{}For simplicity in the presentation, here we have written  bosonic
fields only. Restoring fermion contributions in the formulas is
straightforward and will be done at the end. In what follows we
will omit from most formulas the contribution of the
free coordinates $x^\mu$, which are
decoupled and can be treated as in standard free superstring theory.
After T-duality in the $\vp_1 $ coordinate 
to a new coordinate $\td \vp_1$, and a shift $\vp_2\to\vp_2+b\td
\vp_1$, the Lagrangian becomes
\bea
L &=& \pa _+ r_1\pa _- r_1+r_1^{-2}\pa _+\td \vp_1\pa _-\td \vp_1 +
\pa _+ r_2\pa _- r_2+r_2^2(\pa _+\vp_2+b\pa _+ \td \vp_1) (\pa _-\vp_2+b\pa _- \td
\vp_1)\nn\\
&+& {\cal R}(\phi_0- \ha \log r_1^2)\ ,
\label{bbb}
\eea
where
$$
{\cal R}={1\over 4}\a' \sqrt{g} R^{(2)}\ ,
$$
and $\tilde \vp_1 $ has period $2\pi\alpha '$.
Finally, by performing a T-duality  back in $\td \varphi_1$,
one gets the Lagrangian
\bea
L &=& \pa _+ r_1\pa _- r_1+ \pa _+ r_2\pa _- r_2
+   F (r_1^2 \pa _+ \vp_1\pa _- \vp_1 + r_2^2 \pa _+\vp_2\pa _-\vp_2) 
\nonumber \\
&+&
b F\, r_1^2r_2^2(\pa _+\vp_2\pa _- \vp_1 - \pa _+\vp_1\pa _- \vp_2)
+{\cal R}(\phi_0 + \ha \log F)\ ,
\label{bba}
\eea
$$
F\equiv (1+b^2 r_1^2r_2^2)^{-1}\ .
$$
{}This describes strings propagating in the supergravity background
\bea
ds^ 2 & =& dr_1^2+ dr_2^2+ F(r_1^2d\vp_1^2 +r_2^2d\vp_2^2) \ ,
\nonumber\\
B_{12} & =& {b\, r_1^2r_2^2 F}\ ,\qquad e^{2\phi}=e^{2\phi_0}F\ .
\label{gre}
\eea
By construction, this background is a solution of the string equations to all
orders in $\alpha '$.
The reason is that the model is equivalent to (\ref{bbb}) as a CFT, being related by T-duality,
and (\ref{bbb}) is locally equivalent 
to a background related to flat space by T-duality.

In order to determine the physical string spectrum, one can either consider
the Lagrangian (\ref{bbb}) or (\ref{bba}), since they are equivalent as CFT.   
We shall follow closely ref. \cite{RT}, 
section 5, where a class of string models
obtained by T-duality and shifts are solved,
since there are similarities in the structure of the solution.

In general, 
the solution to the string equations of motion for two T-dual $\sigma$
models are related by $(G_{\mu\nu} +B_{\mu\nu})\pa_\pm x^\nu=\mp
\pa_\pm\tilde x_\mu$. Using this relation, we find the general solution 
to the string equations of motion in the curved background (\ref{gre}),
\be
\vp_1={1\over 2i}\log{X_1\over X_1^*} + b \td\vp_2
\ ,\ \ \ \vp_2={1\over 2i}\log{X_2\over X_2^*}-b \td\vp_1\ ,
\label{khh}
\ee
where
$$
X_1=X_{1+}(\s_+ )+X_{1-}(\s_-)\ ,\ \ \ X_2=X_{2+}(\s_+ )+X_{2-}(\s_-)\ ,
$$
\be
\pa _\pm \td\vp_i=\pm {i\over 2}\Big( X_{i}^*\pa _\pm 
X_{i} -  X_{i} \pa _\pm X_{i}^* \Big)\ .
\ee
Hence
$$
\td\vp_i=2\pi\a'\Big(J_{i-}(\s_-)- J_{i+}(\s_+)\Big) +{i\over 2}(
X_{i+}X_{i-}^*- X_{i-}X_{i+}^*),\ \ \ \ i=1,2\ ,
$$
\be
J_{i\pm} (\s_\pm )\equiv {i\over 4\pi\a' }\int_0^{\s_\pm }d\s_\pm (X_{i\pm}\pa_\pm
X^*_{i\pm}- X^*_{i\pm}\pa_\pm X_{i\pm })\ .
\label{jjjj}
\ee
Using 
\bea
\vp_i(\s+\pi ,\tau )&=& \vp_i(\s,\tau )+2\pi n\ ,
\nonumber\\
\td \vp_i (\s+\pi,\tau ) &=& \td \vp_i(\s,\tau )-2\pi\a' J_i\ ,\ \ \ \ J=J_{iL}+J_{iR}\ ,
\label{klop}
\eea
$$
J_{iL}=J_{i+}(\pi )\ ,\ \ \ \ \ J_R=J_{i-}(\pi )\ ,
$$
one finds that the free fields $X_1\ ,\ X_2$ satisfy the twisted
boundary conditions
\be 
X_1(\s+\pi, \tau  )= e^{ 2\pi i\gd_1} X_1(\s ,\tau )\ ,\qquad
X_2(\s+\pi ,\tau )= e^{ 2\pi i\gd_2} X_2(\s ,\tau )\ ,
\label{reww}
\ee
with 
\be
\gd_1= \a' b J_2\ ,\ \ \ \ 
\gd_2= -\a' b J_1\ .
\ee
Note that $\gd_1, \gd_2 $ are defined modulo $n$, $n=$ integer.
These boundary conditions are satisfied by writing
\be
X_{i\pm} =e^{\pm 2i\gd_i\s_\pm }\chi_{i\pm }\ ,\ \ \ \ \chi_i(\s+\pi ,\tau
)=\chi_i(\s ,\tau )\ .
\label{rew}
\ee
The fields $\chi_{i\pm}$ are single-valued and can be expanded as follows
\be
\chi_{i-}=i\sqrt{\a'\over 2}\sum_n a_{ni} e^{-2in\s_-}\ ,\ \ \ \ \ 
\chi_{i+}=i\sqrt{\a'\over 2}\sum_n \td a_{ni} e^{-2in\s_+}\ .
\label{exde}
\ee

In terms of the free fields $X_1,\ X_2$, the energy-momentum tensor of the string
model (\ref{bba}) is simply
\be
T_{\pm\pm}=\pa _\pm X_1\pa _\pm X_1^*+\pa _\pm X_2\pa _\pm X_2^*\ .
\label{bggg}
\ee
This can be checked by plugging (see eq. (\ref{khh})~)
\be
X_i=r_i e^{i\phi_i}\ ,\ \ \ \ \ \phi_i=\vp_i - b
\epsilon_{ij}\td\vp_j\ .
\ee 
In this way we recover the $T_{\pm\pm}$ in terms of $\vp_1 ,\ \vp_2$
that follows directly 
from the original lagrangian (\ref{bba}), 
\be
T_{\pm\pm}=\sum_{i=1}^2 \big[\pa _\pm r_i\pa _\pm r_i+
{r_i^2\over 1+b^2r_1^2r_2^2}
\pa _\pm \vp_i\pa _\pm \vp_i\big]\ .
\ee
Inserting (\ref{rew}) in (\ref{bggg}), 
the energy-momentum tensor components $T_{\pm\pm}$ take
the form
\be
T_{++}=\sum_{i=1}^2 \big[ 
\pa _+\chi_{i+}\pa _+\chi^*_{i+} + 2i\gd_i (\chi_{i+}\pa _+\chi^*_{i+} -
 \chi_{i+}^*\pa _+\chi_{i+})+4\gd^2_i  \chi_{i+}^*\chi_{i+}\big]\ ,
\ee
\be
T_{--}=\sum_{i=1}^2 \big[ 
\pa _-\chi_{i-}\pa _-\chi^*_{i-} - 2i\gd_i (\chi_{i-}\pa _-\chi^*_{i-} -
 \chi_{i-}^*\pa _-\chi_{i-})+4\gd^2_i  \chi_{i-}^*\chi_{i-}\big]\ .
\ee
Inserting the expansions
(\ref{exde}) and integrating over $\sigma $,
we find the Virasoro operators
\be
L_0={1\over 2} \sum_{i=1}^2  \sum_n (n+\gd_i )^2a_{ni}^* a_{ni} 
\ ,\qquad
\td L_0={1\over 2}\sum_{i=1}^2 \sum_n (n-\gd_i )^2
\td a_{ni}^* \td a_{ni} \ .
\ee
We will also need the expression of the angular momentum components
which are conjugate to $\phi_1 $ and $\phi_2$. In terms of the
mode operators, they are given by
\be
J_{iR}=- {1\over 2} \sum_n (n+\gd_i )a_{ni}^* a_{ni} \ ,
\qquad
J_{iL}=- {1\over 2} \sum_n (n-\gd_i )\td a_{ni}^* \td a_{ni} \ .
\ee

Let us now consider the operator quantization of the model.
Canonical commutation relations for $x_i\equiv r_ie^{i\vp_i}$ imply
\be
[P_{X_i}(\s,\tau ),X^*_j(\s ',\tau)]=-i\delta_{ij} \delta(\s-\s ')\ .
\ee
Hence
\be
[a_{ni},a_{mj}^*]=2(n+ \gd_i )^{-1}\delta_{ij} \delta_{nm}\ ,
\ \ \ \ 
[\td a_{ni},\td a_{mj}^*]=2(n- \gd_i )^{-1}\delta_{ij}\delta_{nm}\ .
\ee
We now introduce standard creation and annihilation operators $b_{n\pm}, \ 
b_{n\pm}^\dagger $, satisfyng $[b,b^\dagger]=1$ by a proper rescaling
of $a_n,\ \td a_n$ as in \cite{RT},
\bea
b_{n+} &=& a_{-n}^*\omega_-\ ,\ \ \ b_{n-}=a_n \omega_+\ ,\  
\nonumber\\
\td b_{n+} &=& \td a_{-n}^*\omega_+\ ,\ \ \ \td b_{n-}=\td a_n
\omega_-\ ,\qquad b_0=\sqrt{\gd/2}\, a_0 \ ,\ \ \td b_0=\sqrt{\gd/2}\, \td a_0^* \ , 
\nonumber\\
\omega_{\pm }&\equiv & \sqrt{\ha (n\pm \gd )}\ ,\ \ \ \ n=1,2,...\ ,\qquad
0<\gd <1\ ,
\eea
where indices 1 and 2 have been omitted.
The Virasoro operators then take the form
\bea
L_0 &=& {1\over 4}\a' p^2_\mu +(\hat N_R- \gd_1 \hat J_{1R}- \gd_2
\hat J_{2R})\ ,
\nonumber\\
\td L_0 &=&
{1\over 4}\a' p^2_\mu +(\hat N_L+ \gd_1\hat J_{1L}+ \gd_2\hat J_{2L})\ ,
\eea
where  $\hat J_{iR},\, \hat J_{iL}$ are given by
\be
\hat J_{R} =J_R-\ha= -b^\dagger_{0} b_{0} -\ha +
\sum_{n=1}^\infty ( b_{n+}^\dagger b_{n+}-
b_{n-}^\dagger b_{n-}\big)+ K_{R}\ ,
\ee
\be
\hat J_{L} =J_L+\ha= \tilde b^\dagger_{0} \td b_{0} +\ha +
\sum_{n=1}^\infty ( \tilde b_{n+}^\dagger \tilde b_{n+}-
\tilde b_{n-}^\dagger \tilde b_{n-}\big)+  K_{L}\ ,
\ee
\be
K_{R}^{\rm (NS)}=-\sum_{r=1/2}^\infty (c_{r}^* c_{r}+c_{-r} c_{-r}^*)\ ,
\ \ \  
K_R^{\rm (R)}=-[d_{0}^*,d_{0}] +
\sum_{n=1}^\infty (d_{n}^* d_{n}+d_{-n} d_{-n}^*)\ ,
\nonumber
\ee
and there is an index $i=1,2$ in all mode operators that has been
omitted for the sake of clarity.
The expression for $ K_{L}$ is similar, with tildes in the mode
operators. We have restored the fermion
contributions, following the notation of \cite{magnetic}
($c_r$ and $d_n$  are the fermion mode operators in the NS and R
sector, respectively).
The eigenvalues of $\hat J_{L,R}$ are
\be
\hat J_{L,R}=\pm (l_{L,R}+\ha )+S_{L,R}\ ,\qquad
\hat J=\hat J_L+\hat J_R=l_L-l_R+S_L+S_R\ ,
\label{polk}
\ee
where $l_L,\, l_R=0,1,2,...$ are Landau quantum numbers 
and the spin $S_L+S_R$ is an integer in the 
NS-NS and R-R sectors, and half-integer in the NS-R and R-NS sectors.

The operators $ \hat N_R,\ \hat N_L$ have the standard expression in
terms
of free creation and annihilation operators,
 $ \hat N_{R,L}=N_{R,L}-a$, with $a^{\rm (NS)}=1/2$,  $a^{\rm
 (R)}=0$, where e.g. in the Ramond sector,
\be
N_R = \sum_{n=1}^\infty n \left(
b_{ni+}^{\dagger} b_{ni+} + b_{ni-}^\dagger
b_{ni-}
+a_{n\alpha }^\dagger a_{n\alpha } +d^*_{ni} d_{ni} +d_{-ni}d^*_{-ni}+
d_{-n\alpha }d_{n\alpha }\right) \ ,
\ee
where summation over $i$ is understood and $a_{n\alpha },\,
d_{n\alpha }$ stand for the remaining ($\alpha=1,...,4$) transverse mode 
operators. For physical states satisfying the GSO condition, the
eigenvalues
are $\hat N_{L,R}=0,1,2,... $~.
The Hamiltonian and level matching constraints are
\be
L_0+\td L_0=0\ ,\ \ \ \ L_0=\td L_0\ .
\ee
They lead to the  string spectrum:
\be
\a' M^2= 2(\hat N_R-\hat \gd_1 \hat J_{1R}-\hat \gd_2 \hat J_{2R})+
2(\hat N_L+\hat \gd_1 \hat J_{1L}+\hat \gd_2 \hat J_{2L})\ ,
\label{zzz}
\ee
\be
\hat N_R=\hat N_L \ ,
\label{lmc}
\ee
where $\hat \gd_i=\gd_i-[\gd_i]$, and
\be
\gd_1= \a' b(\hat  J_{2R}+ \hat J_{2L}) \ ,\ \ \ \ 
\gd_2=- \a' b(  \hat J_{1R}+ \hat J_{1L})\ .
\ee
Note that $ \a' b $ is dimensionless.

\medskip

A few remarks are in order:

\begin{itemize}

\item The spectrum has a periodic dependence on the twist parameters
$\gd_i $. This must be the case, since the boundary conditions 
(\ref{rew}) are unchanged if we replace $\gd_i\to\gd_i +n_i$, $n_i=$
integer. Due to the presence of fermions, the actual periodicity is
$\gd_i\to\gd_i +2n_i$.
When $2n_i\leq \gd_i<2n_i+1$, $i=1,2$ one should use the standard GSO projection.
When one of the $\gd_i $ is in the interval $2n_i+1\leq \gd_i<2n_i+2$, one 
should use the reversed GSO projection, meaning that only states having half-integer eigenvalues of
the operators $\hat N_{L,R}$ will survive, i.e. $\hat N_{L,R}=-\ha, \ha, {3\over 2},...$
(see \cite{magnetic}). 


\item The fact that in (\ref{zzz}) $\hat \gd_i \hat J_{iL}$ appear with the opposite sign 
of $\hat \gd_i \hat J_{iR}$ is due to our conventions. What is independent
of  conventions is that the terms proportional 
to the Landau numbers $l_{iL}$ and
$l_{iR}$ both contribute with positive sign to $M^2$, and this is
of course the case in eq.~(\ref{zzz}).

\item If $\a' b $ is irrational, then $\gd_1,\, \gd_2$ are not integer numbers
for any $\hat J_1,\hat J_2\neq 0$. When  $\a' b $ is rational, $\a'
b=p/q$, there are sectors  $\hat J_{1}$ or $\hat J_2=qn$,
where one of the $\gd_i $ is an integer and the corresponding
$\hat\gd_i $ vanishes.

\item If one of the $\hat \gd_i $ vanishes, 
say $\hat\gd_2$, then
the zero mode structure in the plane 2 changes. The oscillator modes 
$b_{02},b_{02}^\dagger$ and $\tilde b_{02},\tilde b_{02}^\dagger$ 
(giving rise to Landau numbers $l_{2R}, \, l_{2L}$) are replaced by
$x_{2},\,  x_2^*, \, p_2,\,  p_2^*$.

\item If $\hat \nu_1$ and $\hat\nu_2$ do not vanish, 
then $M^2\geq 0$ for any $b$
(see below).

\end{itemize}

As an application, let us consider states of minimal energy for a given level $N\equiv \hat N_R=\hat N_L$.
From the explicit representation in terms of creation and annihilation operators, one can see that
the $S_{iR},\ S_{iL}$ satisfy the bounds
\be
\big|S_{1R}\pm S_{2R}\big|\leq \hat N_R+1\ ,\qquad
\big|S_{1L}\pm S_{2L}\big|\leq \hat N_L+1\ .
\label{mabo}
\ee
We consider a state having $l_{iL}=l_{iR}=0$, and
$$
S_{1R}+S_{2R}=N+1\ ,\qquad S_{1L}+S_{2L}=-N-1\ ,
$$
It follows that $\hat J_{1}=-\hat J_2$. We assume $\hat J_2=S_{2L}+S_{2R}>0$ and $0<\a' b<1$.
Then $\gd_1=\gd_2=\a' b\, s,\ s\equiv S_{2L}+S_{2R}$. The mass formula takes the form
\be
\a' M^2=4N\big(1-(\gd_1 -[\gd_1])\big)\ .
\label{vfe}
\ee
This is manifiestly positive definite.
It is possible to take a limit, which is similar to the limits 
studied in \cite{russo},
where most string states decouple. The number of surviving states at each level is proportional to $N$.
In the present case, we consider $\a' b=1-\vep$. Then 
$$
\gd_1 -[\gd_1]= \a' b\, s- [\a' b\, s]  =1-\vep s\ , 
$$
where we have assumed $\vep s<1$, which is a valid assumption for any given $s$,
since we are going to take the limit $\vep\to 0$.
Now write $\a'=\vep \a'_{\rm eff}$, and take the limit $\vep\to 0$
with fixed $\a'_{\rm eff}$. In this limit, the masses of all states with
$S_{1R}+S_{2R}<N+1$ or $ S_{1L}+S_{2L}>-N-1$ go to infinity.
For the special states considered above, 
the mass formula (\ref{vfe}) takes the form
\be
\a'_{\rm eff} M^2=4N\, s\ ,\qquad   0<s\leq 2N+2\ ,
\label{vfez}
\ee
Thus these states have finite mass after the limit  $\vep\to 0$. 
Note that one can also consider a limit with $\a' b=p/q-\vep $, where
there are surviving states.

\section{String model II: TsT on three polar planes}

\subsection{One-parameter deformation}

The starting point is the string theory Lagrangian  
\be
L=\sum_{i=1}^3 \left( \pa _+ r_i\pa _- r_i
+r_i^2\pa _+\phi_i\pa _- \phi_i \right)\ ,
\label{hhzz}
\ee
or, in Cartesian coordinates,
\be
L=\sum_{i=1}^3  \pa _+ X_i\pa _- X_i^*\ ,
\label{yyy}
\ee
where
\be
\phi_1=\psi-\vp_1'\ ,\ \ \ \phi_2=\psi -\vp_2'\ ,\ \ \ \
\phi_3=\psi+\vp_1'+\vp_2'\ .
\label{defi}
\ee
Here we omit other free coordinates in the string theory Lagrangian
as well as fermion contributions. They will be incorporated later.

Now we proceed as in the model of section 2, by performing a T-duality 
transformation in
the $\vp_1 '$ variable to a new variable $\td \vp_1$, and
a shift: $\vp_2'=\vp_2+b\td \vp_1$.
After T-duality in $\td \vp_1$ to the T-dual  variable $\vp_1$,
one obtains a  
final  Lagrangian which is symmetric in $\vp_1 $, $\vp_2$,
representing a curved string background with B-field components and dilaton.
This model is constructed in the appendix A of \cite{LM}.

Using the relation $(G_{\mu\nu} +B_{\mu\nu})\pa_\pm x^\nu=\mp
\pa_\pm \tilde x_\mu$ between 
the solutions to the string equations of motion for two T-dual $\sigma$
models, we find  the solution
\bea
\vp_1 &=& \vp_1' +b\td\vp_2= 
{1\over 3} (\phi_2+\phi_3-2\phi_1)+b \td \vp_2\ ,
\nn \\
\vp_2 &=& \vp_2' -b\td\vp_1= {1\over 3} (\phi_1+\phi_3-2\phi_2)-b \td \vp_1\ ,
\nn \\
\psi &=& {1\over 3} (\phi_1+\phi_2+\phi_3)\ ,
\label{defiz}
\eea
where
\bea
\pa_\pm \td\vp_1 &=& \mp \left( r_3^2 \pa_\pm \phi_3 - 
r_1^2\pa_\pm \phi_1 \right) \ ,
\nonumber\\
\pa_\pm \td\vp_2 &=& \mp \left( r_3^2 \pa_\pm \phi_3 - 
r_2^2\pa_\pm \phi_2 \right) \ .
\eea
Using  (\ref{defi}), (\ref{defiz}) and
the fact that $\vp_1,\ \vp_2$ and $\psi $ are $2\pi $ periodic, 
we find 
the boundary conditions of $\phi_i$ variables, 
\bea
\phi_1(\s +\pi )&=& \phi_1(\s ) +b \Delta \td \vp_2\ ,\ \ \ \nn \\
\phi_2(\s +\pi )&=& \phi_2(\s ) -b \Delta \td \vp_1\ ,\ \ \ \nn \\
\phi_3(\s +\pi )&=& \phi_3(\s ) - b \Delta \td \vp_2+b \Delta \td \vp_1
\ ,\
\eea
with
\be
\Delta \td \vp_1=2\pi\a' (J_1-J_3)\ ,\ \ \ \ \ \Delta \td \vp_2=2\pi\a' (J_2-J_3)\ ,
\ee
$$
J_i=J_{iL}+J_{iR}\ ,\ \ \ \ J_{iL}=J_{i+}(\pi )\ ,\ \ \ 
J_{iR}=J_{i-}(\pi )\ .
$$
The operators $J_i$ are as in (\ref{jjjj}), with $i=1,2,3$.
 
Thus the twists $\gd_i$ (defined as in (\ref{reww})~) in the free
fields $X_i$ are given by 
\be
\gd_1=\a' b (J_2-J_3)\ ,\ \ \ \gd_2=\a' b (J_3-J_1)\ ,\ \ \ \ 
\gd_3=\a' b (J_1-J_2)\ .
\ee
%
We then proceed exactly as in section 2: 
we redefine $X_i$ in terms of single-valued fields $\chi_i$ as in
(\ref{rew}), and the
the expressions that follow are the same as in section 2,
with the only difference that now $i=1,2,3$.

Therefore, we find the string spectrum
\be
\a' M^2= 2(\hat N_R-\hat \gd_1 \hat J_{1R}- \hat \gd_2 \hat J_{2R}-\hat \gd_3 \hat J_{3R} )+
2(\hat N_L+\hat \gd_1 \hat J_{1L}+\hat \gd_2 \hat J_{2L}+ \hat \gd_3 \hat J_{3L})\ ,
\label{xxx}
\ee
\be
\hat N_R=\hat N_L\ .
\label{jjj}
\ee
We recall the notation $\hat\gd_i=\gd_i-[\gd_i]$,
$\hat J_{iR}=J_{iR}-\ha $, $\hat J_{iL}=J_{iL}+\ha $, so that 
$\hat J_i=\hat J_{iL}+\hat J_{iR}=J_i$.
The same remarks given at the end of section 2 apply to this model.


\subsection {Three independent deformations}

Here we consider three independent deformations $b_i $,
which are the analog to the 3-parameter deformation 
of $AdS_5\times S_5$ studied in \cite{frolov,frt2}.
This model is obtained by a sequence of transformations, 
$({\rm TsT})_{b_1}({\rm TsT})_{b_2}({\rm TsT})_{b_3}$.
Following the same procedure as in the previous sections, we now find the spectrum
\be
\a' M^2= 2\Big( \hat N_R+\hat N_L-\hat \gd_1 (\hat J_{1R}-\hat J_{1L})
- \hat \gd_2( \hat J_{2R}-\hat J_{2L})
-\hat \gd_3 (\hat J_{3R}-\hat J_{3L})\Big)\ ,
\label{xxxa}
\ee
\be
\hat N_R=\hat N_L\ ,
\label{jjja}
\ee
where $\hat \gd_i=\gd_i-[\gd_i]$,
\be
\gd_1=  \a' (b_3 \hat J_2-b_2\hat J_3)\ ,\ \ \ 
\gd_2=  \a'(b_1 \hat J_3-b_3 \hat J_1)\ ,\ \ \ \ 
\gd_3=  \a'(b_2\hat J_1-b_1\hat J_2)\ ,
\label{fas}
\ee
or
\be
\gd_i=\a'  \epsilon_{ijk} b_k \hat J_j\ .
\ee
In the case $b_1=b_2=b_3=b$, we recover the mass spectrum (\ref{xxx})
of the model of section 3.1.

\medskip

For generic values of $b_1,\, b_2,\, b_3$, all supersymmetries are broken.
An important question is whether the mass spectrum contains tachyons.
To look for a tachyon, we shall consider a state with $ \hat \gd_1,\, \hat \gd_2,\, 
\hat \gd_3 $ different from zero, and with maximum value of $ (\hat J_{1R}-\hat J_{1L})
$.

\smallskip

The different situations that typically arise are illustrated below by
considering different regions of the parameter space.

\medskip

\noindent 1) All $\gd_i $ are in the interval $0<\gd_i<1 $:

\smallskip

{} The $S_{iR},\ S_{iL}$ satisfy the bounds
\be
\big|S_{1R}\pm S_{2R}\pm S_{3R}\big|\leq \hat N_R+1\ ,\qquad
\big|S_{1L}\pm S_{2L}\pm S_{3L}\big|\leq \hat N_L+1\ .
\ee
{}So we choose
$S_{1R}=N+1$, $S_{1L}=-N-1$, where $N\equiv \hat N_R=\hat N_L$,
$l_{1L}=l_{1R}=0$
and, in addition,
\be
l_{2R}=1\ ,\ \ l_{2L}=S_{2R}=S_{2L}=0\ ,\ \ 
\label{kmmm}
\ee
\be
l_{3L}=1\ ,\ \ l_{3R}=S_{3R}=S_{3L}=0\ .\ \ 
\label{knnn}
\ee
Hence we have (see (\ref{polk})~)
\be
\hat J_{1R}=N+\ha\ ,\ \ \ \ \hat J_{1L}=- N -\ha\ ,\ 
\ee
\be
\hat J_{2R}=-{3\over 2}\ ,\ \ \ \ \hat J_{2L}=\ha\ ,\ \ \qquad
\hat J_{3R}=-\ha\ ,\ \ \ \ \hat J_{3L}= {3\over 2}\ .\ 
\label{kbbb}
\ee
For the deformation parameters, we assume $-1<b_2+b_3<0$, 
$0<b_1<1$. 
Note that this range already 
excludes the supersymmetric case $b_1=b_2=b_3$.
In what follows we set $\a' =1$.
{} From eq. (\ref{fas}), we find
\be
\hat\gd_1= \big| b_2+b_3 \big|\ ,\qquad \hat \gd_2=b_1\ ,\qquad \hat\gd_3=b_1\ .
\ee
Then the mass formula takes the form
\be
 M^2=2 \Big( 2N +4b_1 - \big| b_2+b_3
\big|\  (2N+1)\Big)\ .
\label{uuuu}
\ee
These states become tachyonic for
\be
\big| b_2 +b_3 \big|> b_{\rm cr}\ ,\qquad   b_{\rm
cr}\equiv {2N +4b_1\over 2N+1}\ .
\label{yyyyy}
\ee 
Thus,  for any $N=0,1,2,...$, the states are tachyonic for sufficiently large $\big| b_2+b_3 \big|$.
The assumption $\big| b_2+b_3 \big|<1$ in the tachyonic regime
(\ref{yyyyy}) is satisfied by choosing $4b_1<1$.
This leaves a wide range of $b$-parameters where these tachyons exist.

Finally, one can show that fermions have positive mass squared, as expected.

\medskip

\noindent 2) $\gd_1 $ is in the interval $-1 <\gd_1<0 $:

\smallskip

As pointed out in section 2, in this case the GSO projection is the reversed one.
Now $\hat N_{L,R}$ take the values $\hat N_{L,R}=-\ha,\ha,{3\over 2},...$ and we have the bound 
\be
\big|S_{1R}\pm S_{2R}\pm S_{3R}\big|\leq \hat N_R+\ha\ ,\qquad
\big|S_{1L}\pm S_{2L}\pm S_{3L}\big|\leq \hat N_L+\ha\ .
\ee
{} We choose
$S_{1R}=N+\ha $, $S_{1L}=-N-\ha$,  $N\equiv \hat N_R=\hat N_L$, 
$l_{1L}=l_{1R}=0$,
so that
\be
\hat J_{1R}=N\ ,\ \ \ \ \hat J_{1L}=- N \ .\ 
\ee
In the planes 2 and 3, we choose the same quantum numbers as in
eqs.~(\ref{kmmm}), (\ref{knnn}), (\ref{kbbb}).

For the deformation parameters, we now 
assume $0<b_2+b_3<1$, \  $0<b_1<1$. 
{} From (\ref{fas}), we now find
\be
\hat\gd_1= 1- b_2 -b_3 \ ,\qquad \hat \gd_2=b_1\ ,\qquad \hat\gd_3=b_1\ .
\ee
Then the mass formula (\ref{xxxa}) becomes
\be
 M^2= 2 \Big( 2N +4b_1 - (1- b_2 -b_3 )\  2N\Big)= 
2\Big( 4b_1 + ( b_2+b_3 )\  2N\Big)\ .
\ee
The state $N=-\ha $ is tachyonic for
\be
 b_2+b_3 > 4 b_1\ .
\label{uio}
\ee 
In the supersymmetric case, $b_1=b_2=b_3$, the condition 
(\ref{uio}) is not satisfied and the state has a positive squared mass.

\medskip

\noindent 3) $\gd_1 $ is in the interval $-2 <\gd_1<-1 $:

\smallskip

Here we have again the standard GSO projection.
We choose exactly the same quantum numbers as in the case 1).
But now we assume $1<b_2+b_3<2$,  $0<b_1<1$. 
The supersymmetric case $b_1=b_2=b_3$ is included in the discussion, and it is interesting
to see how tachyons disappear.
{} We have
\be
\hat\gd_1= 2- b_2-b_3 \ ,\qquad \hat \gd_2=b_1\ ,\qquad \hat\gd_3=b_1\ .
\ee
Then the mass formula takes the form
\be
 M^2=2 \Big( 2N +4b_1 - (2- b_2-b_3)\  (2N+1)\Big)\ .
\label{moon}
\ee
These states are tachyonic for
\be
2-b_2-b_3>{2N+4b_1\over 2N+1}\ ,\qquad 4b_1< 1\ .
\ee
Now let us specialize to the supersymmetric case $b_1=b_2=b_3$. 
Then $4b_1=2(b_2+b_3)=4-2\hat \gd_1$. Hence
\be
 M^2=2\Big( 2N +(4-2\hat\gd_1) - \hat\gd_1\  (2N+1)\Big)=
2 \Big( (2N +4)(1 - \hat\gd_1) +\hat\gd_1 \Big)\ ,
\ee
which is positive definite, since, by definition, $0<\hat\gd_1<1$.

\section{Energies of short strings in the Lunin-Maldacena\\ deformation of $AdS_5\times S^5$  }


Given the parallel between the models considered here
and the Lunin-Maldacena deformation $(AdS_5\times
S^5)_\beta $, an interesting question is whether there may be
a relation between the corresponding string spectra. 
For strings of size much less than the $AdS_5$ or $S^5$ radius $R$,
the string dynamics is essentially as in flat spacetime.
We expect that the spectrum of such short strings in the
Lunin-Maldacena background $(AdS_5\times
S^5)_\beta $ will have a similar structure as the spectra discussed in
this paper,
with the change $\alpha'\to \alpha'/\sqrt{\lambda }$,
and $\a' b\to \hat \beta/\sqrt{\lambda }$, 
$\hat \beta\equiv \beta \sqrt{\lambda }$.
The parameter $\beta $ is assumed to be real and it is what appears in 
the Yang-Mills superpotential
${\rm Tr}\big[ e^{i\pi\beta} \Phi_1\Phi_2\Phi_3-e^{-i\pi\beta}
\Phi_1\Phi_3\Phi_2\big]$. Taking a flat-space limit of the string
spectrum in  $(AdS_5\times S^5)_\beta $
 necessarily requires sitting
at some point of $S^5$ (e.g. $r_3=1,\, r_1=r_2=0$),
and considering short strings. This procedure breaks the $Z_3$ symmetry 
associated with exchange of the 1-2-3 planes. The resulting string spectrum
approaches a truncation of the spectrum of the model of section 3.1
(the spectrum of short strings in $(AdS_5\times S^5)_\beta $
cannot be described by the full spectrum (\ref{xxx}),  because the latter involves
oscillator modes associated with six dimensions, as opposed to the
 five dimensions of $S^5$).
One interesting problem would be to compare the string spectrum (\ref{xxx})
with the energy of semiclassical  short strings in $(AdS_5\times
 S^5)_\beta $ having $1\ll \hat  J_i\ll\sqrt{\lambda }$.


The existence of tachyons in the three-parameter model of section 3.2 
suggests that there could also be tachyons in the analog model of 
\cite{frolov,frt2}.
It will be difficult to
see such possible tachyons in a semiclassical approximation of large $N$.  
In particular, note that for the existence of the above tachyon states,
 it is essential that there is a  ``1" in $2N+1$ in (\ref{uuuu}) and
 in
(\ref{moon}).
The origin of this 1 is a normal ordering contribution, and it is
negligible in a semiclassical approximation where $N\gg 1$. 
There are tachyons at any given string level number $N$ 
in  some regions of parameter space, including low values of $N$, in particular $N=0$.
{}From the point of view of the dual gauge theory, low $N$ means 
short operators (the string level number $N$ should not be confused of course with N
of U(N)).
It would be interesting to see if there is a
counterpart of the tachyon instabilities in the dual
non-supersymmetric gauge theory
of \cite{frolov,frt2}.
It would also be interesting to 
understand the limit 
taken at the end of section 2 (related to
$\beta\to p/q$) within the ${\cal N}=1$ superconformal gauge theory.

\smallskip

\section*{Acknowledgments}

I would like to thank A. Adams, J. Maldacena, T. Mateos, M. Spradlin and especially A.A. Tseytlin
 for  useful discussions and comments.
This work is
supported in part by the European
EC-RTN network MRTN-CT-2004-005104, and by MCYT FPA
2004-04582-C02-01 and CIRIT GC 2001SGR-00065.

\setcounter{section}{0}
\setcounter{subsection}{0}

\setcounter{equation}{0}



\begin{thebibliography}{20}

\bibitem{RT}
  J.~G.~Russo and A.~A.~Tseytlin,
  ``Exactly solvable string models of curved space-time backgrounds,''
  Nucl.\ Phys.\ B {\bf 449}, 91 (1995)
  [arXiv:hep-th/9502038].



\bibitem{magnetic}
  J.~G.~Russo and A.~A.~Tseytlin,
  ``Magnetic flux tube models in superstring theory,''
  Nucl.\ Phys.\ B {\bf 461}, 131 (1996)
  [arXiv:hep-th/9508068].


\bi{taka} 
 T.~Takayanagi and T.~Uesugi,
  ``Orbifolds as Melvin geometry,''
  JHEP {\bf 0112}, 004 (2001)
  [arXiv:hep-th/0110099].


\bi{flux} J.~G.~Russo and A.~A.~Tseytlin,
  ``Supersymmetric fluxbrane intersections and closed string tachyons,''
  JHEP {\bf 0111}, 065 (2001)
  [arXiv:hep-th/0110107].



\bi{taka2}  T.~Takayanagi and T.~Uesugi,
  ``D-branes in Melvin background,''
  JHEP {\bf 0111}, 036 (2001)
  [arXiv:hep-th/0110200].




\bi{costa} M.~S.~Costa and M.~Gutperle,
  ``The Kaluza-Klein Melvin solution in M-theory,''
  JHEP {\bf 0103}, 027 (2001)
  [arXiv:hep-th/0012072].
 
\bi{RT01} J.~G.~Russo and A.~A.~Tseytlin,
  ``Magnetic backgrounds and tachyonic instabilities in closed superstring
  theory and M-theory,''
  Nucl.\ Phys.\ B {\bf 611}, 93 (2001)
  [arXiv:hep-th/0104238].


\bi{blau} M.~Blau, M.~O'Loughlin, G.~Papadopoulos and A.~A.~Tseytlin,
  ``Solvable models of strings in homogeneous plane wave backgrounds,''
  Nucl.\ Phys.\ B {\bf 673}, 57 (2003)
  [arXiv:hep-th/0304198].


\bi{hashi} A.~Hashimoto and L.~Pando Zayas,
  ``Correspondence principle for black holes in plane waves,''
  JHEP {\bf 0403}, 014 (2004)
  [arXiv:hep-th/0401197].


\bi{fiol}  N.~Drukker, B.~Fiol and J.~Simon,
  ``Goedel-type universes and the Landau problem,''
  JCAP {\bf 0410}, 012 (2004)
  [arXiv:hep-th/0309199].



\bi{LM}
  O.~Lunin and J.~Maldacena,
  ``Deforming field theories with U(1) x U(1) global symmetry and
    their gravity duals,''
  hep-th/0502086.

\bi{frt}
S.~A.~Frolov, R.~Roiban and A.~A.~Tseytlin,
  ``Gauge - string duality for superconformal deformations of N = 4 super
  Yang-Mills theory,''
  hep-th/0503192.

\bibitem{LEST}
  R.~G.~Leigh and M.~J.~Strassler,
  ``Exactly marginal operators and duality in four-dimensional N=1
  supersymmetric gauge theory,''
  Nucl.\ Phys.\ B {\bf 447}, 95 (1995)
  [hep-th/9503121].



\bibitem{NIPR}
  V.~Niarchos and N.~Prezas,
  ``BMN operators for N = 1 superconformal Yang-Mills theories and
    associated string backgrounds,''
  JHEP {\bf 0306}, 015 (2003)
  [hep-th/0212111].



\bi{frolov}
S.~Frolov,
  ``Lax pair for strings in Lunin-Maldacena background,''
  JHEP {\bf 0505}, 069 (2005)
  [hep-th/0503201].

\bi{nunez} 
U.~Gursoy and C.~Nunez,
 ``Dipole deformations of N = 1 SYM and supergravity backgrounds with U(1) x
 U(1) global symmetry,''
  arXiv:hep-th/0505100.





\bibitem{beisert}
  N.~Beisert and R.~Roiban,
  ``Beauty and the twist: The Bethe ansatz for twisted N = 4 SYM,''
  arXiv:hep-th/0505187.


\bi{koch}
R.~de Mello Koch, J.~Murugan, J.~Smolic and M.~Smolic,
  ``Deformed PP-waves from the Lunin-Maldacena background,''
  hep-th/0505227.



\bi{mat}
T.~Mateos,
  ``Marginal deformation of N = 4 SYM and Penrose limits with continuum
  spectrum,''
  hep-th/0505243.




\bibitem{bobev}
  N.~P.~Bobev, H.~Dimov and R.~C.~Rashkov,
  ``Semiclassical strings in Lunin-Maldacena background,''
  arXiv:hep-th/0506063.



\bibitem{freedman}
  D.~Z.~Freedman and U.~Gursoy,
  ``Comments on the beta-deformed N = 4 SYM theory,''
  arXiv:hep-th/0506128.




\bibitem{penati}
  S.~Penati, A.~Santambrogio and D.~Zanon,
 ``Two-point correlators in the beta-deformed N = 4 SYM at 
the next-to-leading order,''
  arXiv:hep-th/0506150;
  A.~Mauri, S.~Penati, A.~Santambrogio and D.~Zanon,
  ``Exact results in planar N = 1 superconformal Yang-Mills theory,''
  arXiv:hep-th/0507282.



\bibitem{frt2}
  S.~A.~Frolov, R.~Roiban and A.~A.~Tseytlin,
  ``Gauge-string duality for (non)supersymmetric deformations of N = 4 super
  Yang-Mills theory,''
  JHEP {\bf 0507}, 045 (2005)
  [arXiv:hep-th/0507021].

\bibitem{Ahn}
  C.~Ahn and J.~F.~Vazquez-Poritz,
  ``Deformations of flows from type IIB supergravity,''
  arXiv:hep-th/0508075.


\bibitem{kuzenko}
  S.~M.~Kuzenko and A.~A.~Tseytlin,
  ``Effective action of beta-deformed N=4 SYM theory and AdS/CFT,''
  arXiv:hep-th/0508098.




\bibitem{russo}  J.~G.~Russo,
  ``Strong magnetic limit of string theory,''
  JHEP {\bf 0506}, 005 (2005)
  [arXiv:hep-th/0504187].


\end{thebibliography}
\end{document}